\documentclass[twocolumn, showpacs, longbibliography, prapplied,aps, amsmath, amssymb,superscriptaddress]{revtex4-1}
\usepackage{amsfonts} 
\usepackage{graphicx} 
\usepackage[colorlinks,linkcolor=blue,citecolor=blue,urlcolor=blue]{hyperref}
\usepackage{float}
\usepackage[normalem]{ulem}
\usepackage[switch,columnwise]{lineno}

\begin{document}

\title{\Large{Enhancing the speed and sensitivity of a nonlinear optical sensor with noise}}

\author{Said R. K. Rodriguez} \email{s.rodriguez@amolf.nl}

\affiliation {Center for Nanophotonics, AMOLF, Science Park 104, 1098 XG Amsterdam, The Netherlands}

\date{\today}

\begin{abstract}

We demonstrate how noise can be turned into an advantage for optical sensing using a nonlinear cavity.  The cavity is driven by a continuous wave laser into the regime of optical bistability. Due to the influence of fluctuations, the cavity randomly switches between two states. By analyzing residence times in these two states, perturbations to the resonance frequency of the cavity can be detected. Here, such an analysis is presented as a function of the strength of the perturbation and of the noise.  By increasing the standard deviation of the noise, we find that the detection speed  increases monotonically while the sensitivity peaks at a finite value of the noise strength. Furthermore, we discuss how noise-assisted sensing can be optimized in state-of-the-art experimental platforms, relying solely on the minimum amount of noise present in the cavity due to its dissipation. These results open new perspectives for the ultrafast detection of nanoparticles, contaminants, gases, or other perturbations to the resonance frequency of an optical resonator, at low powers and in noisy environments.
\end{abstract}
\narrowtext

\maketitle

\section{Introduction}
A sensor is a device that reports a change in its environment. The sensor-environment coupling leads to dissipation which, according to the fluctuation-dissipation theorem~\cite{Kubo66}, makes the output of the sensor necessarily noisy.  This minimum amount of noise places a lower bound on the magnitude of the perturbation that a linear dissipative sensor can detect within a certain time.   Additional noise in the sensor or the environment typically degrades the sensing performance further; the measurement time needed to detect a certain perturbation only increases with the noise strength.

In 2002, Gammaitoni and Bulsara introduced a sensor whose performance can be enhanced by nonlinearity and noise; they called it a noise activated nonlinear dynamical sensor (NANDS)~\cite{Gammaitoni02}. The physics of a NANDS is reminiscent of the Brownian particle in a double-well potential (DWP) mastered by Kramers~\cite{Kramers}. If the DWP is symmetric, the average residence time of the particle in each well is the same. However, if the DWP is tilted, the residence time difference (RTD) is non-zero on average. Thus, RTD measurements can be used to detect perturbations affecting the symmetry of the potential. A similar sensing scheme can be realized with noisy nonlinear oscillators. A cubic nonlinearity leads to an effective DWP, and noise can make the oscillator switch between two states corresponding to the minima of the DWP~\cite{Gammaitoni98}.

Until now, the RTD sensing scheme has been successfully employed in the context of magnetic field detection \cite{Ando05, Baglio11}.  Experiments and calculations on NANDS have focused on configurations involving a periodic modulation of the DWP~\cite{Ando05, Baglio11, Bulsara03, Nikitin03, Dari10, Nikitin13},  where noise plays a secondary role with respect to the periodic force.  This is likely the best detection strategy in systems where slow dynamics and weak noise make fully noise-activated sensing too slow or impractical. As shown ahead, the situation is different for several technologically relevant optical systems.

Linear optical resonators are already well-known for their sensing capabilities, mainly attributed to their high quality factors, small mode volumes, high operation frequencies, and the possibility to easily and remotely read-out their state with light~\cite{vanDuyne03, Vollmer08, Zhu10,  Offermans11, Zijlstra12, Shao13, Vollmer15, Rosenblum15, Kelkar15, Zhi17}.  Nonetheless, as the size and power budget of optical sensors continue to decrease, noise is playing an increasingly deleterious role in their performance. In this vein, many efforts have focused on realizing nanophotonic sensors with enhanced sensitivities,  as expected near exceptional points for example~\cite{Wiersig14, Chen17, Zhong19}. However, as long as those sensors remain linear, time-invariant, and passive, noise stands on the way of exploiting the enhanced sensitivity to detect small perturbations~\cite{Langbein18, Lau18, Mortensen18}.   These developments suggest that a detection strategy harnessing rather than avoiding noise, as in the RTD scheme, may lead to a new frontier in optical sensing.

Here we demonstrate how the minimum amount of noise present in a nonlinear optical resonator, as dictated by the fluctuation-dissipation theorem, can be turned into an advantage for sensing. The sensing scheme we propose is based on measuring the RTD without any periodic driving. This scheme can be realized in  resonators supporting optical bistability~\cite{Gibbs, Lipson04, Notomi05, Mabuchi11,  Deveaud15, Rodriguez17, Fink18}, i.e. two steady-states with different photon number at a single driving condition. We will show that optically bistable resonators can be used as sensors with the following remarkable properties: i) a detection speed that increases monotonically with the standard deviation of the noise and ii) a sensitivity that is maximum for a particular value of the standard deviation of the noise.

\section{A noisy nonlinear optical cavity as a sensor}

\begin{figure}[!]
 \centerline{{\includegraphics[width=\linewidth]{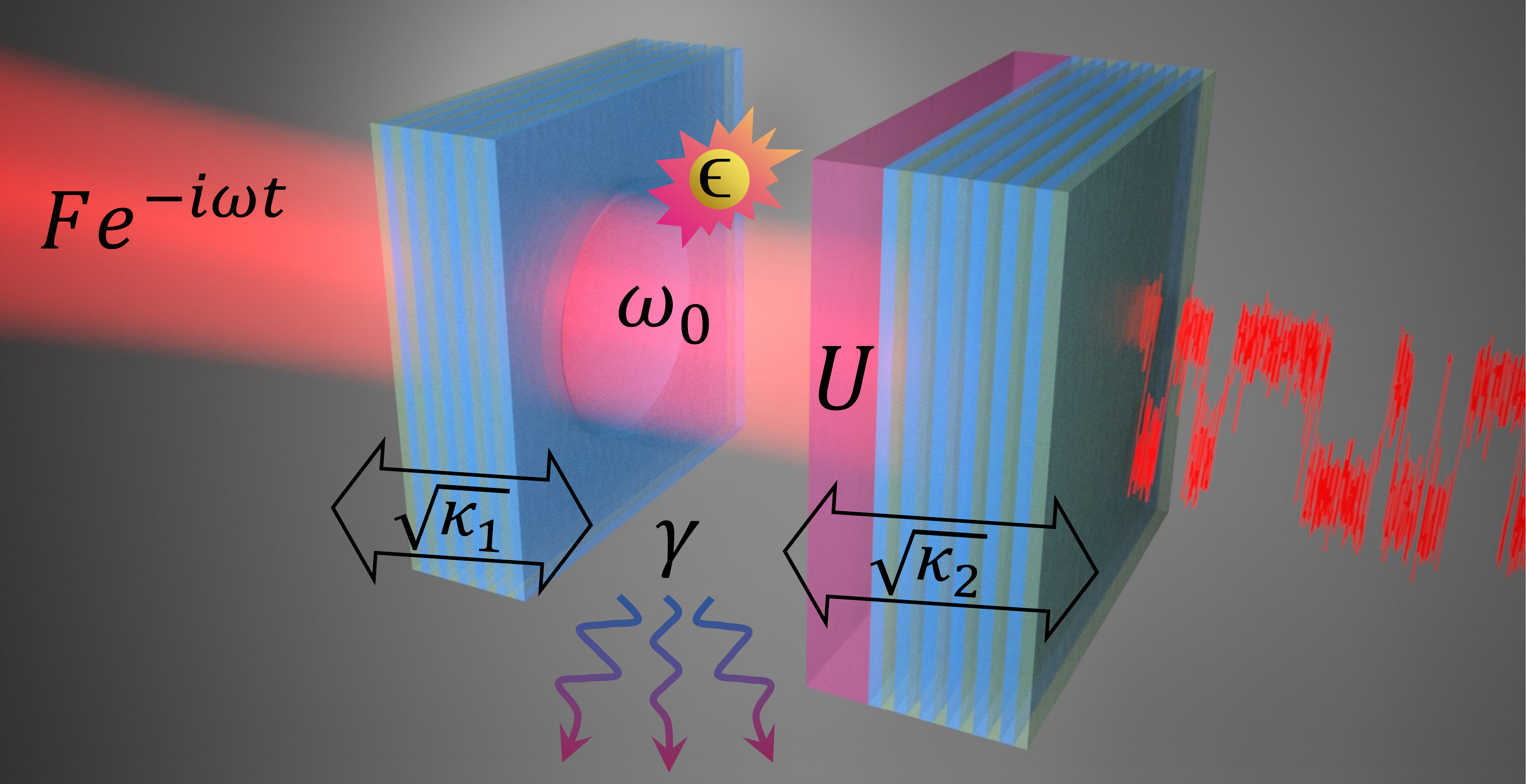}}}\caption{A single mode cavity with resonance frequency $\omega_0$, intrinsic loss rate $\gamma$, and Kerr nonlinearity corresponding to effective photon-photon interactions of strength $U$, is driven by a coherent field of amplitude $F$ and frequency $\omega$. $\kappa_{1,2}$ are the leakage rates of the optical mode through the mirrors.  The transmitted intensity shows that the cavity randomly switches between two states due to the nonlinearity and the influence of fluctuations. The residence time of the cavity in these two states is highly sensitive to perturbations to the resonance frequency. A nanoparticle is an example of such a perturbation, here labeled as $\epsilon$. }
\label{fig1}
\end{figure}

For concreteness but without loss of generality, we consider a single-mode Fabry-P\'{e}rot cavity (see Fig.~\ref{fig1}) as our resonator and sensor. The cavity is made by two distributed Bragg reflectors  facing each other. One of the mirrors is concave, as realized in Refs.~\cite{Trichet14, Trichet16, Bitarafan} for example. The other mirror is planar, and coated with a nonlinear material leading to optical bistability. Single-mode operation is possible when the radius of curvature of the concave mirror and the resonance linewidths are sufficiently small for the transverse cavity modes to be well-isolated from each other. In the following, we will investigate the dynamics of such a single-mode nonlinear cavity when its resonance frequency is perturbed. The perturbation can be a single nanoparticle entering the cavity's mode volume, as investigated in Ref.~\onlinecite{Trichet16} for example.

The cavity is driven by a continuous wave laser with frequency $\omega$ and amplitude $F$. In a frame rotating at the driving frequency, the dynamics of the intra-cavity field $\alpha$ is governed by the following equation

\begin{equation}\label{eq1}
i\dot{\alpha}=\left(-\Delta - i \frac{\Gamma}{2} + U (\vert \alpha \vert^2 -1) \right)\alpha + i \sqrt{\kappa_{1}}F	e^{-i\omega t} + D \xi(t).
\end{equation}

\noindent $\Delta = \omega-\omega_0$ is the detuning between the laser frequency and the cavity resonance frequency $\omega_0$. $\Gamma = \ \gamma + \kappa_1 + \kappa_2$ is the total loss rate, with $\gamma$ the internal cavity loss rate and $\kappa_{1,2}$ the leakage rates of the cavity field across the mirrors. $U$ is the effective photon-photon interaction strength, associated with the nonlinear material inside the cavity. The term $D \xi(t)$ accounts for white noise, with standard deviation $D$, in the two quadratures of the light field. The stochastic term $\xi(t) = \xi(t)' + i \xi(t)''$ is a complex Gaussian processes with zero mean and delta correlated, i.e., $\langle \xi' \rangle = \langle \xi'' \rangle = 0$ and $\langle \xi'(t) \xi'(t+t') \rangle=  \langle \xi''(t) \xi''(t+t') \rangle =  \delta(t')$. Moreover, $\xi'$ and $\xi''$ are mutually delta correlated: $\langle \xi'(t) \xi''(t+t') \rangle = \delta(t')$.

\begin{figure*}[!]
 \centerline{{\includegraphics[width=\linewidth]{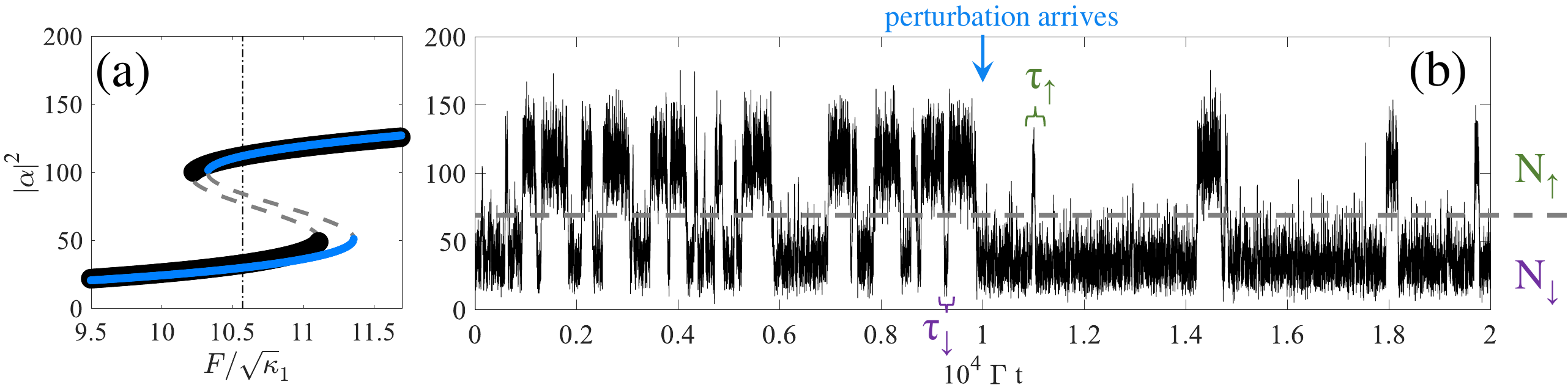}}}\caption{ (a) Steady-state number of photons in the cavity $|\alpha|^2$ as a function of the driving amplitude $F$ referenced to the input leakage rate $\kappa_1$. Solid and dashed curves correspond to stable and unstable states, respectively. Black and blue curves correspond to unperturbed and perturbed cavities. The perturbation is defined by letting the laser-cavity detuning $\Delta \rightarrow \Delta (1+\epsilon)$, with  $\epsilon=0.02$. (b) Time-evolution of $|\alpha|^2$ influenced by noise. Mid-way in the calculation we introduce the aforementioned perturbation $\epsilon=0.02$. The horizontal line at $|\alpha|^2 = 69.8$ indicates the threshold above (resp. below) which the cavity resides in state $N_{\uparrow}$  (resp. $N_{\downarrow}$). Examples of the corresponding residence times are labeled as  $\tau_{\uparrow}$  and $\tau_{\downarrow}$. For all calculations in Figs.~\ref{fig2}(b), ~\ref{fig3}, ~\ref{fig4}, and~\ref{fig5}, the values of the parameters used in the calculations, relative to $\kappa_1$, are: $\kappa_2=2\kappa_1/3$, $\Gamma = 2 \kappa_1$,  $F=10.57 \sqrt{\kappa_1}$ [indicated by the vertical dash-dotted line in (a)], $U/\Gamma = 0.01$, $\Delta/\Gamma=1.0965$, and $D=\sqrt{\Gamma/2}$.  The same values were used for Fig.~\ref{fig2}(a), except that $D=0$. A time step of $0.1 \Gamma^{-1}$ was used for all calculations from  Fig.~\ref{fig2}(b) onwards. }
\label{fig2}
\end{figure*}

Let us briefly review the steady-state response of the cavity without noise, obtained by setting $\dot{\alpha}=0 $ and  $D=0$ in Eq.~\ref{eq1}. The thick black curve is Fig.~\ref{fig2}(a) shows the number of photons in the cavity  $|\alpha|^2$  as  a function of $F$ for a detuning $\Delta/\Gamma=1.0965$. For this detuning the cavity transmission displays a maximum within the bistability~\cite{Rodriguez19}. This is convenient for transmission measurements, but we will not exploit the enhanced transmission in our present analysis. Figure~\ref{fig2} shows the typical `$S$' curve of bistability, observed whenever $\Delta > \sqrt{3} \Gamma/2$ and $U>0$.  In the bistability, the cavity can reside in either of  two steady-states depending on the initial conditions and the driving history of the system.  Since there is no noise, the residence time of the cavity in either state is infinite.

In the presence of noise, the cavity can randomly switch between two states with different average number of photons as expected based on the steady-state analysis. Such a switching behavior has been observed in the transmission of a laser-driven bistable cavity influenced by quantum fluctuations, for example~\cite{Rodriguez17, Fink18}. To illustrate this behavior, we performed stochastic calculations using the xSPDE package~\cite{xSPDE}. In Fig.~\ref{fig2}(b) we plot $|\alpha|^2$  as a function of time for constant $F=10.57 \sqrt{\kappa_1}$, $D=\sqrt{\Gamma/2}$, and for one realization of the noise.  Mid-way in the calculations we introduced a perturbation $\epsilon$ to the resonance frequency of the cavity. More precisely, the perturbation was set by letting $\Delta \rightarrow \Delta (1+\epsilon)$, with $\epsilon=0.02$. Since $\Delta/\Gamma$ is of order one, $\epsilon$ can be approximately interpreted as the fractional change in the resonance frequency relative to the linewidth.  Notice in Fig.~\ref{fig2}(b) how the perturbation influences the switching behavior and biases the system towards the low density state. This biasing can be regarded as a tilting of the effective DWP for the intracavity light field.  For reference, the thin blue  curve in  Fig.~\ref{fig2}(a) shows the steady-state number of photons in the perturbed cavity. The biasing of the system towards the low-density state can be inferred from the enhanced proximity of the driving amplitude [vertical dashed line in Fig.~\ref{fig2}(a)]  to the threshold  value for which the system jumps from the upper to the lower branch.

\begin{figure}[!]
 \centerline{{\includegraphics[width=\linewidth]{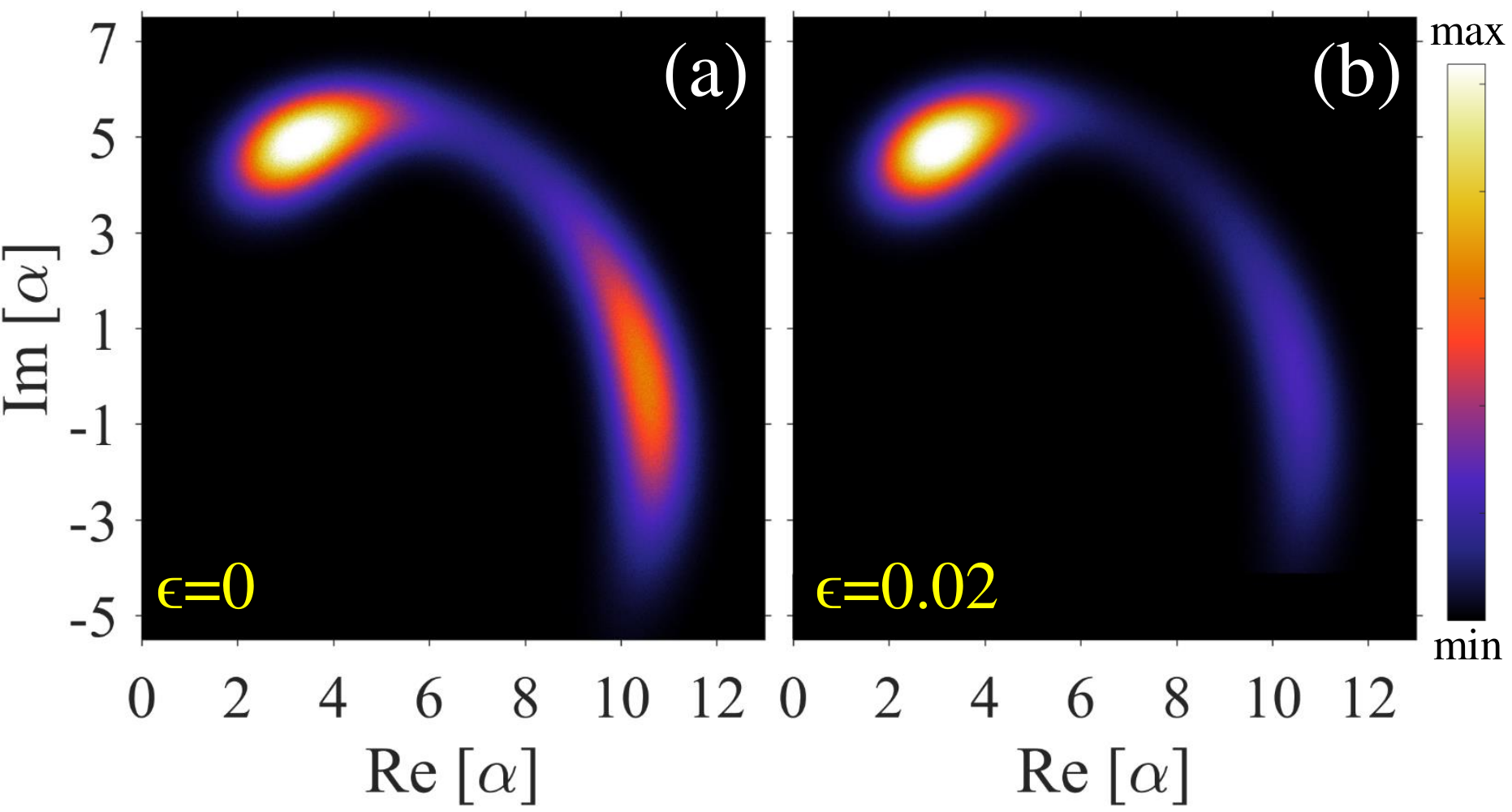}}}\caption{ Probability distribution of the complex field $\alpha$ for the (a) unperturbed and (b) perturbed cavity. The perturbation corresponds to letting $\Delta \rightarrow \Delta (1+\epsilon)$, with $\epsilon=0.02$. Other values of parameters used in the calculations are given in Fig.~\ref{fig2}.}
\label{fig3}
\end{figure}

Similarities between light confined in a nonlinear cavity and a Brownian particle in a DWP can be recognized in the probability distribution of the complex field $\alpha$. To calculate such distribution, we let the system evolve for a long time and plot a histogram of the state of the system as a function of the real and imaginary parts of $\alpha$. By ``long time'' we mean that $t \nu \gg 1$, with $\nu$ the average rate at which the cavity switches between states.  For large photon numbers ($|\alpha|^2 \gg 1$) in the bistability region, such a histogram corresponds to the Wigner function obtained through a quantum approach~\cite{Vogel89, Casteels17}.

The results of our calculations are shown in Fig.~\ref{fig3}(a) for $\epsilon=0$, and in Fig.~\ref{fig3}(b) for $\epsilon=0.02$. The values of the model parameters used for the calculations in Fig.~\ref{fig3}(a) and~\ref{fig3}(b) are the same as in the first and second half of the time evolution in Fig.~\ref{fig2}(b), respectively. Both panels in Fig.~\ref{fig3} display a bimodal distribution indicating bistability. The peaks of these distributions represent the minima of the DWP in Kramers' problem. Note, however, that the effective potential for the light field involves two dynamical variables, namely the real and imaginary parts of $\alpha$. Comparison of Figs.~\ref{fig3}(a) and ~\ref{fig3}(b) reveals that the perturbation decreases the probability of finding the system in the state around Re$[\alpha]$=10.5 and Im$[\alpha]$=0, which is the  high-density state. This is akin to tilting the DWP towards the low-density state. Notice also that the high-density state has a significantly reduced uncertainty along the real component of $\alpha$. This squeezing may be exploited for sensing by performing homodyne detection, but we will not comment on this further.

\section{Sensitivity and detection speed}
We would now like to quantify the sensitivity of the cavity to perturbations in $\Delta$. Such perturbations are common in optical sensing settings such as the one illustrated in Fig.~\ref{fig1}. Note, however, that the stochastic dynamics of the cavity  actually depends on the ratio $\Delta/\Gamma$~\cite{Rodriguez17}.   Here, for simplicity, we shall assume that $\Gamma$ is unaffected by the perturbation. In case such an assumption cannot be made, perturbations that increase both $\Delta$ and $\Gamma$ can be regarded as effectively weaker than those which increase only $\Delta$ or $\Gamma$.

The first step in our sensing protocol is to define a threshold density $N_{th}$ above (resp. below) which the cavity is said to reside in state $N_{\uparrow}$ (resp.  $N_{\downarrow}$). The Appendix  shows how $N_{th}$ can be defined by analyzing the probability density function of $|\alpha|^2$.  Next, we define residence times $\tau_{\uparrow}$ and $\tau_{\downarrow}$  as the time intervals for which the cavity continuously resides in states $N_{\uparrow}$ and $N_{\downarrow}$, respectively. For example,  in Fig.~\ref{fig2}(b) we indicate two events with residence time  $\tau_{\uparrow}$ and  $\tau_{\downarrow}$, and we indicate $N_{th}$ with a horizontal dashed line. Naturally, the random nature of the switching leads to distributions of residence times. Therefore, a reliable detection strategy should be constructed based on the properties of these distributions. In particular, we will inspect the time-averaged residence time difference $\delta \tau =  \tau_{\uparrow} - \tau_{\downarrow}$.

\begin{figure}[!]
 \centerline{{\includegraphics[width=\linewidth]{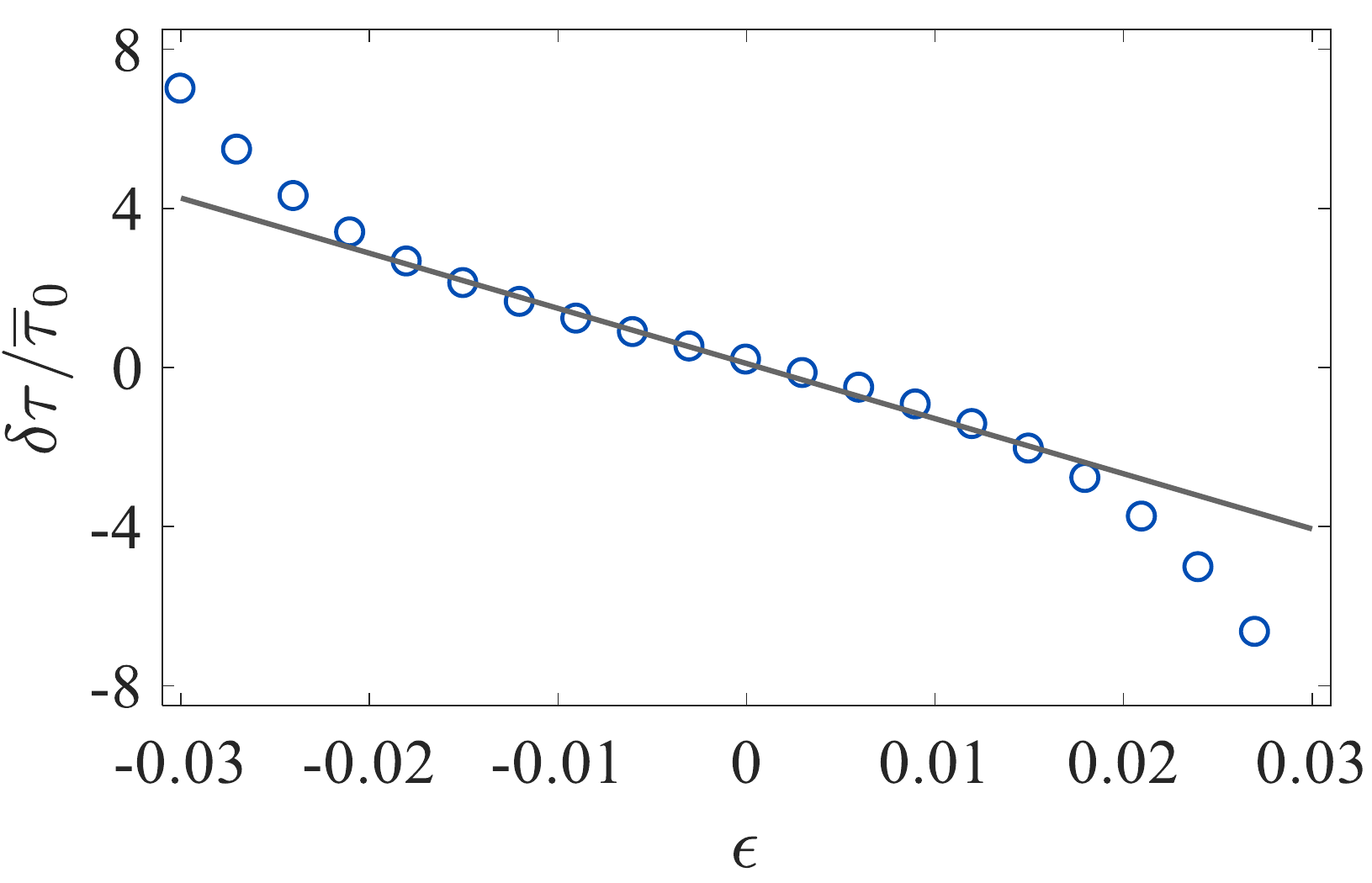}}}\caption{ The average residence time difference $\delta \tau$ between $N_{\uparrow}$ and $N_{\downarrow}$ states (see Fig.~\ref{fig2}) versus the strength of the perturbation $\epsilon$ in the detuning.  $\delta \tau$ is divided by $\overline{\tau}_0 = [\tau_{\uparrow}(\epsilon=0) + \tau_{\downarrow}(\epsilon=0)]/2 $, which is the average residence time of the system in states $N_{\uparrow}$ and $N_{\downarrow}$ at $\epsilon=0$.  In this way, changes in  $\delta \tau$ relative to the residence times of the unperturbed cavity can be estimated. The gray line is a linear fit to the data for small $\epsilon$, from which the sensitivity of the cavity to small perturbations is obtained. Values of parameters used in the calculations are given in Fig.~\ref{fig2}. }
\label{fig4}
\end{figure}

In Fig.~\ref{fig4} we present  calculations of  $\delta \tau$ as a function of $\epsilon$, with otherwise identical driving conditions to those in Fig.~\ref{fig2}(b) and Fig.~\ref{fig3}. For all $\epsilon$, $\delta \tau$ was first time-averaged based on 4000 residence events,  and then ensemble-averaged over 100 realizations with different noise seeds. By varying the noise seed we take into account that nominally identical measurements can give slightly different values of $\delta \tau$ in finite time. Most importantly, by varying the noise seed we avoid having the same realization of the noise for any two perturbations; that would enable the detection of arbitrarily small perturbations because the stochastic trajectories of cavity fields with different perturbations would be correlated.

Figure~\ref{fig4} shows that the average $\delta \tau$  is, in general, a nonlinear function of $\epsilon$. This eases the detection of relatively large perturbations. For small $\epsilon$ the average $\delta \tau$ is approximately linear. This is convenient for calibrating the sensor. From the slope of the fitted line in Fig.~\ref{fig4} we extract a value of the sensitivity to small perturbations $\mathcal{S}= \frac{\partial \delta \tau}{ \partial \epsilon} = 138.6 \pm 9$ s, with the uncertainty corresponding to a $95$\% confidence interval.

\begin{figure}[!]
 \centerline{{\includegraphics[width=\linewidth]{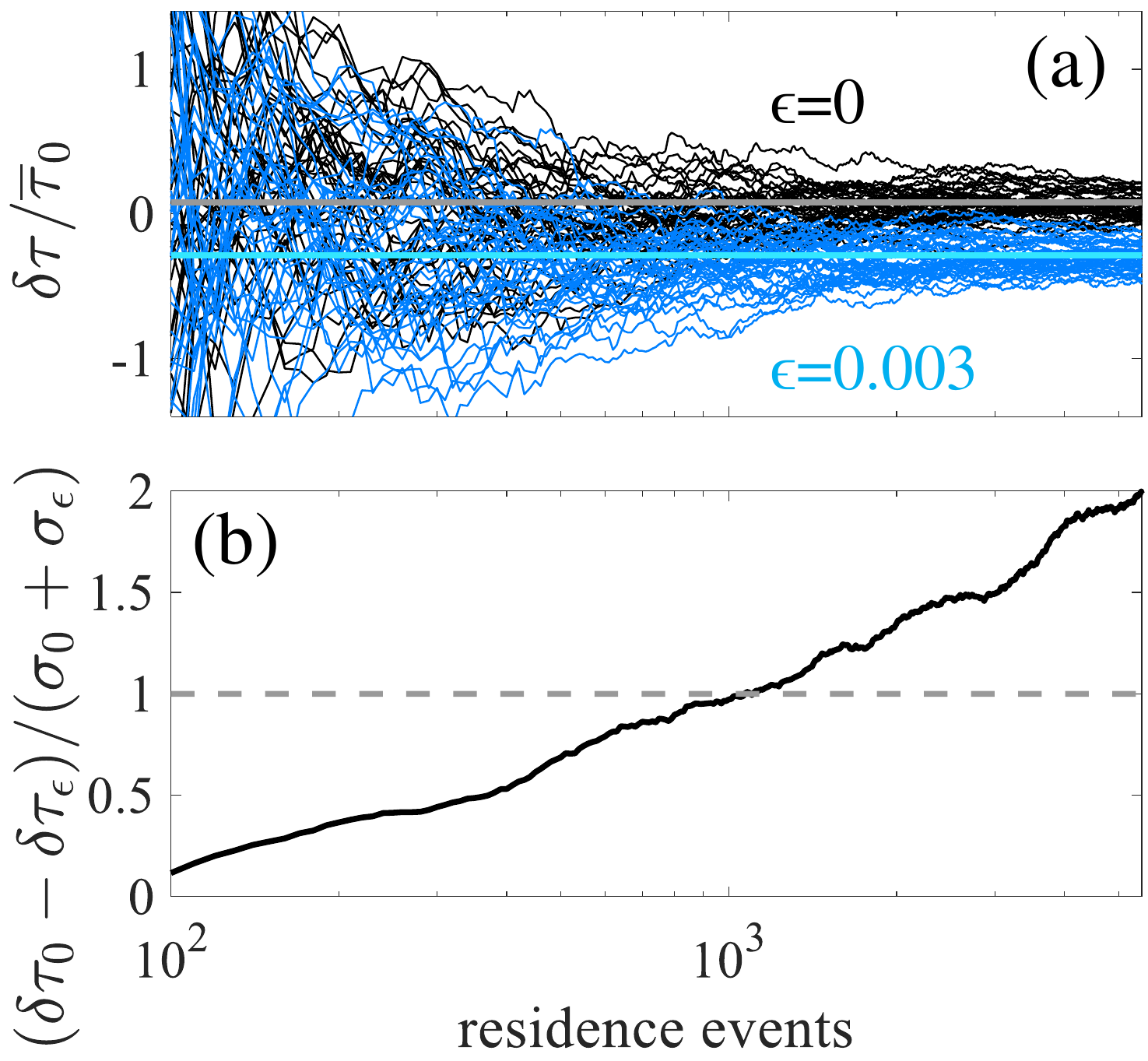}}}\caption{ (a) Average residence time difference, as in Fig.~\ref{fig4}, as a function of the number of residence events over which the averaging is performed. Black and blue lines correspond to unperturbed and perturbed cavities, respectively. Different lines of the same color correspond to different realizations of the noise. The thick horizontal gray and blue lines indicate the long-time average residence time difference for the unperturbed and perturbed cavity, respectively. (b)  $\delta \tau_0$ and $\delta \tau_{\epsilon}$ are the average residence time difference without and with the perturbation, respectively. $\sigma_0$ and  $\sigma_{\epsilon}$ are standard deviations of the residence time difference distributions without and with the perturbation, respectively. Each element in a residence time difference distribution is associated with a different noise seed in the calculation. $(\delta \tau_0 - \delta \tau_{\epsilon})/(\sigma_0 + \sigma_{\epsilon}) >1$, indicated by the dashed gray line, can be considered as the detection threshold. Values of parameters used in the calculations are given in Fig.~\ref{fig2}.}
\label{fig5}
\end{figure}

Another important figure of merit is the detection speed. Assessing the detection speed of our stochastic nonlinear cavity is tantamount to answering the following question: how many residence events are needed to detect a certain perturbation? In Fig.~\ref{fig5} we illustrate how this question can be answered, taking a tiny perturbation of $\epsilon=0.003$ as an example. Figure~\ref{fig5}(a) shows several  calculations of $\delta \tau$ as the number of residence events involved in the  time-averaging  increases.  Black and blue lines correspond to an unperturbed ($\epsilon=0$)  and  perturbed ($\epsilon=0.003$) cavities, respectively. Various lines of the same color correspond to nominally identical configurations, but with different realizations of the noise $\xi(t)$ obtained from different seeds. The results in Fig.~\ref{fig5}(a) show how, as the number of residence events involved in the time-averaging increases,  the spread in the values of $\delta \tau$  decreases.  For $t \to \infty$, all values of $\delta \tau$ for a given $\epsilon$ converge to a single value indicated by a thick horizontal line. The long-time difference in $\delta \tau$ with and without perturbation  is determined by the aforementioned sensitivity.

The number of residence events needed to detect $\epsilon=0.003$ with a single noise realization can be determined by comparing two quantities: i) the change in the average RTD due to the perturbation, i.e. $\delta \tau_0 - \delta \tau_{\epsilon}$,  and ii) the sum of the standard deviations of the RTD distributions, i.e. $\sigma + \sigma_{\epsilon}$. We recall that each element of the RTD distributions is associated with a different noise seed. A simple criterion for detection can be $|\delta \tau_0 - \delta \tau_{\epsilon}| > (\sigma_0 + \sigma_{\epsilon})$. This corresponds to a shift in $\delta \tau$ that is greater than the sum of the uncertainties. Using this criterion, Fig.~\ref{fig5}(b) shows that $\epsilon=0.003$ can be detected with $\sim1000$ residence events or more. Note that we have taken a stringent criterion for the detection threshold. A reliable detection strategy can still be constructed with significantly less residence events, provided that we accept larger probabilities of false alarm and missed detection~\cite{KayDT}.

\begin{figure}[!]
 \centerline{{\includegraphics[width=\linewidth]{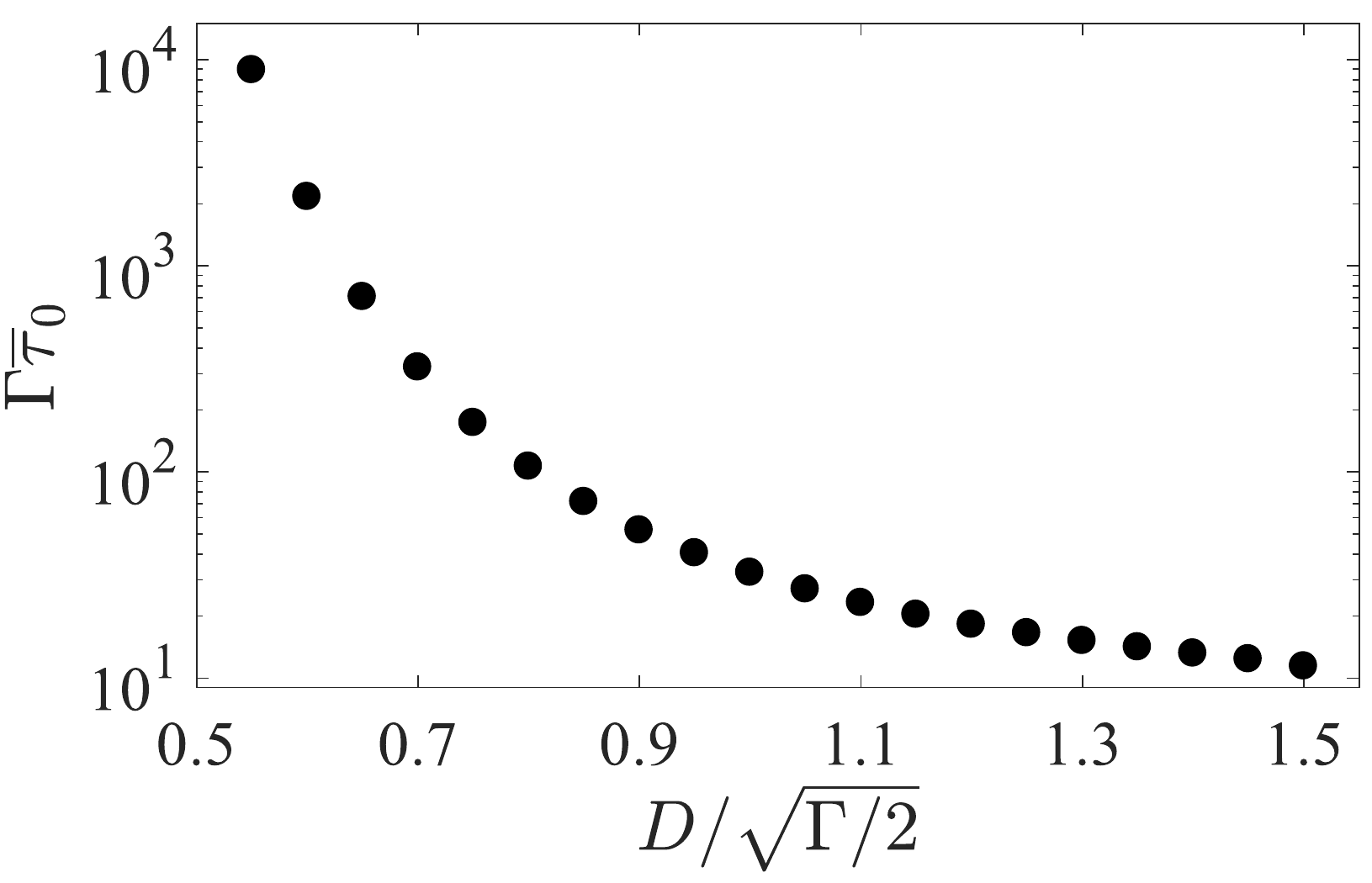}}}\caption{Average residence time for $\epsilon=0$, $\overline{\tau}_0$, times the total loss rate $\Gamma$, as a function of the standard deviation of the noise $D$ referenced to $\sqrt{\Gamma /2}$.}
\label{fig6}
\end{figure}

Within a fixed measurement time, the number of residence events  increases with the standard  deviation of the noise $D$. To illustrate this, in Fig.~\ref{fig6} we plot the average residence time of the unperturbed cavity in the two states, $\overline{\tau}_0 = [\tau_{\uparrow}(\epsilon=0) + \tau_{\downarrow}(\epsilon=0)]/2$ as a function of $D/\sqrt{\Gamma/2}$. Notice that $\overline{\tau}_0$ decreases by $\sim3$ orders of magnitude within a factor of 2.5 increase in  $D/\sqrt{\Gamma/2}$. This result,  together with the result in Fig.~\ref{fig5}(b), seems to suggest that the sensing performance of our nonlinear cavity improves indefinitely with the addition of noise. However, this is not the case because the sensitivity also depends on the noise.

\begin{figure}[!]
 \centerline{{\includegraphics[width=\linewidth]{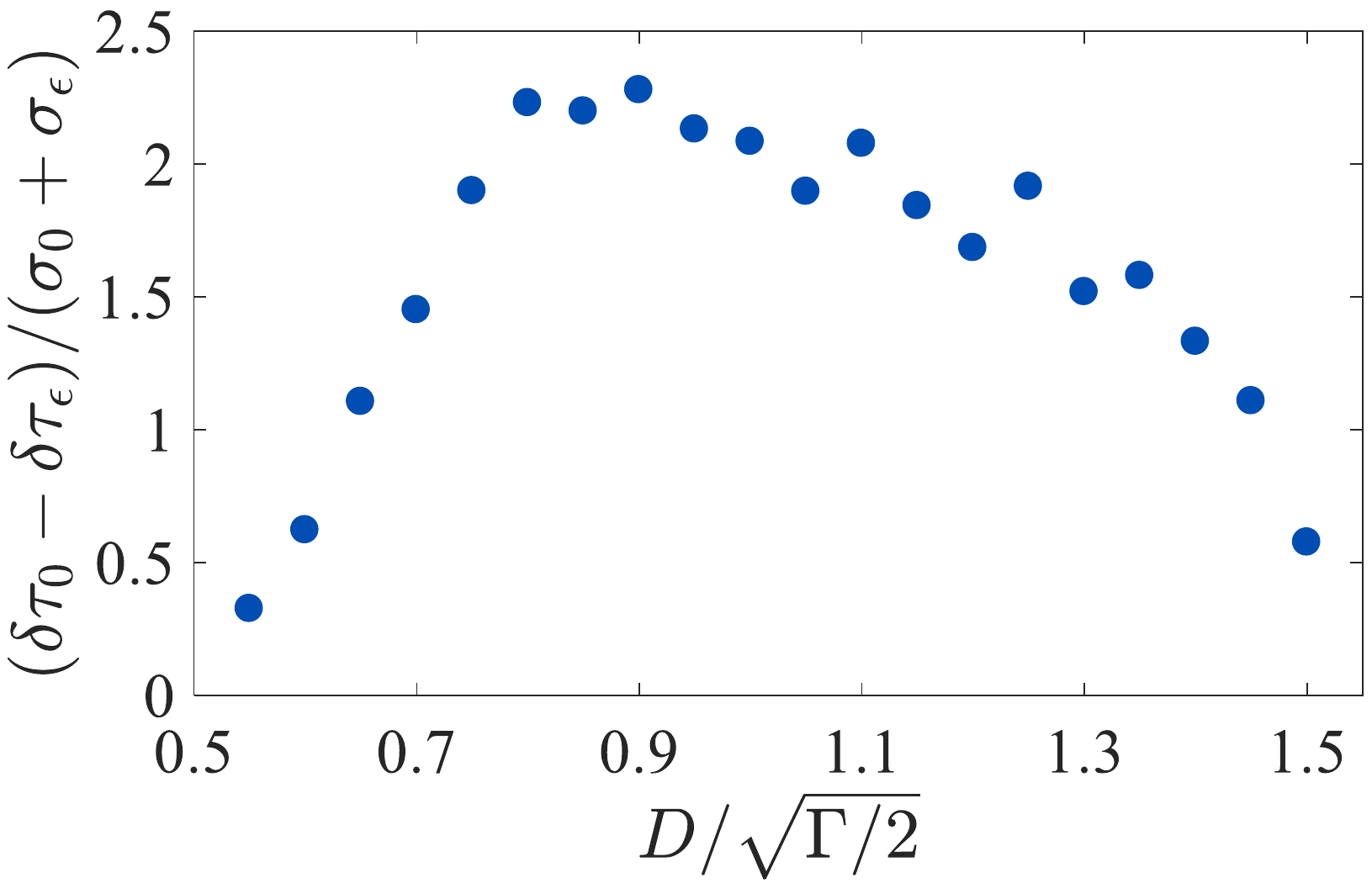}}}\caption{Same as in Fig.~\ref{fig5}(b), but now as a function of $D\sqrt{\Gamma /2}$. Other values of parameters used in the calculations are the same as in  Fig.~\ref{fig2}.}
\label{fig7}
\end{figure}

In Fig.~\ref{fig7} we plot $(\delta \tau_0 - \delta \tau_{\epsilon}) /   (\sigma_0 + \sigma_{\epsilon})$ as a function of $D/\sqrt{\Gamma/2}$, for a fixed measurement time $2\times 10^5 \Gamma^{-1}$. As before, we have taken $\epsilon=0.003$ as an example. Recall that $|\delta \tau_0 - \delta \tau_{\epsilon}|$ is proportional to the sensitivity, and $|\delta \tau_0 - \delta \tau_{\epsilon}| >  (\sigma_0 + \sigma_{\epsilon})$ is the detection threshold we previously defined. According to the results in Fig.~\ref{fig7}, $\epsilon=0.003$ can only be detected within a finite range of non-zero noise. For the values of the parameters we have chosen (achievable with modern semiconductor cavities~\cite{Deveaud15, Rodriguez17, Fink18}, for example) the sensitivity is greatest for $D \approx 0.9 \sqrt{\Gamma/2}$. $D = \sqrt{\Gamma/2}$ is exactly the minimum amount of noise in the cavity demanded by the fluctuation-dissipation theorem. Moreover, for  $D = \sqrt{\Gamma/2}$ the average residence time  at the center of the bistability,  where $\tau_{\uparrow} = \tau_{\downarrow}$, corresponds  to the so-called quantum tunneling time of bistability~\cite{Rodriguez17}. Details on how to calculate this tunneling time based on a quantum master equation approach can be found in References~\onlinecite{Risken87, Casteels17}.

To understand why the sensitivity peaks for a finite amount of noise, consider the behavior of the system for  extreme values of $D$. For $D \to 0$, the number of residence events within the measurement time decreases. Consequently, $\sigma + \sigma_{\epsilon}$ increases and the presence of a small perturbation becomes increasingly uncertain.  Conversely, for large $D$ the residence time becomes too short.  Effectively, this can be associated with a potential barrier that is too small. Such a small barrier makes it practically impossible to detect perturbations affecting the symmetry of the potential. In particular, the sensitivity is degraded when the height of the barrier is much less than: i) the change in energy between the minima of the DWP due to the perturbation, and ii) the average energy in the fluctuations. Hence, a finite amount of noise is needed for a detection strategy based on RTDs to succeed.

Practically, a system can be operated with the optimum amount of noise by either injecting noise, or by judiciously selecting the laser intensity and the laser-cavity detuning. The laser power and detuning determine the number of photons involved in the bistability, which in turn determines the average residence times. Thus, by varying the laser parameters  one can effectively adjust the potential barrier. Note also that the overall optimum amount of noise needs not to coincide with the peak sensitivity. For instance, the results in Fig.~\ref{fig6} and Fig.~\ref{fig7} show that  increasing $D/\sqrt{\Gamma/2}$ from 0.8 to 1.2 degrades the RTD shift by $\sim7\%$, while the the number of residence events that can be acquired within a fixed measurement time increases by $\sim460\%$. Correspondingly, the amount of noise optimizing the overall performance of the sensor is above the value for which the sensitivity peaks.

In order to connect our proposal with a potential experimental realization, we would like to specify parameter values that may be achieved using present-day technologies.  For example, consider a cavity with a resonance frequency $\omega_0=360$ THz, a total loss rate $\Gamma=10$ GHz, and $U/\Gamma=0.01$; these values are typical for III-V semiconductor cavities in Ref.~\cite{Rodriguez17, Deveaud17, Fink18, Lagoudakis18}. The calculations in Figs.~\ref{fig2}-~\ref{fig7} were all done for $U/\Gamma=0.01$, so we can immediately read out the corresponding time scales. In particular, in Fig.~\ref{fig5} we noted that a perturbation of $\epsilon=0.003$ can be detected with $\sim1000$ residence events, using the minimum amount of noise present in the cavity due to the dissipation.  For a switching rate of $10 \gamma$, which can be achieved in the range $0.9 \gtrsim \Delta/\Gamma \gtrsim 1$, we conclude that a tiny shift in the resonance frequency that is  $\sim 0.3\%$  of the linewidth can be detected within a measurement time of $\sim 10$ ns. Larger perturbations (still on the order of a few percent of the resonance linewidth) that make the system depart out of the bistability can be detected with fewer residence events, since the system will stop switching. For the parameter values we have considered, this means that perturbations of a few percent of the resonance linewidth can be detected within $\sim 0.1$ ns.

In our analysis so far, we have not mentioned the fact that residence times also depend on the strength of photon-photon interactions relative to the dissipation, i.e., $U/\Gamma$~\cite{Rodriguez17}. As $U/\Gamma$  decreases, the number of photons involved in the bistability increases. Consequently, residence times for fixed $\Delta/\Gamma$ will be longer at the center of the bistability. This is not a problem for our sensing scheme, as long as greater laser power is available and the frequency of the laser or the cavity resonance can be tuned. As shown in Ref.~\onlinecite{Rodriguez17}, for small detunings ($\Delta/\Gamma \sim 0.9$) residence times at the center of the bistability are all on the order of $10 \Gamma^{-1}$ for vastly different values of $U/\Gamma$. Thus, our sensing scheme can be realized in systems with vastly different, albeit finite, nonlinearity. The validity of our model is only expected to break down for extremely strong nonlinearity, i.e. $U/\Gamma\sim1$. In those cases, our mean-field equation plus stochastic terms (the so-called truncated Wigner approximation~\cite{CarusottoRMP}) is no longer valid, and a full quantum master equation approach is needed.

\section{Conclusion}
In conclusion, we have proposed a noisy nonlinear optical cavity as a reliable and ultrafast sensor. Such a sensor can be used to detect perturbations to the resonance frequency of the cavity. The detection speed of this sensor increases with the noise strength, while its sensitivity peaks for a particular noise strength. This unusual dependence of the sensing performance on noise may open new possibilities for sensing at low optical powers and in noisy environments.  Our sensor can  be used to detect contaminants or gases affecting the resonance frequency of an optical resonator, for example. While for the sake of concreteness we have focused the discussion on a Fabry-P\'{e}rot cavity, our results hold for any single-mode nonlinear optical resonator, which could be a microdisk~\cite{Vollmer15}, a ring resonator~\cite{Lipson04}, or a photonic crystal cavity~\cite{Notomi05}. In some of these systems,  bistability emerges from a  nonlinear optical response which is of thermal origin.  In those cases, a similar sensing scheme can still be realized but the thermal relaxation time will limit the maximum operation speed of the sensor. Alternatively, an overdamped optically levitated nanoparticle~\cite{Ricci} can also be used as a noise-assisted sensor in a similar way.

\section*{Acknowledgements}
The author is grateful to Javier del Pino for critical comments, and to Zhou Geng and Kevin Peters for stimulating discussions. This work is part of the research programme of the Netherlands Organisation for Scientific Research (NWO). The author acknowledges financial support from NWO through a Veni grant with file number 016.Veni.189.039.

\section*{Appendix: Definition of $N_{th}$}
\begin{figure}[!]
 \centerline{{\includegraphics[width=\linewidth]{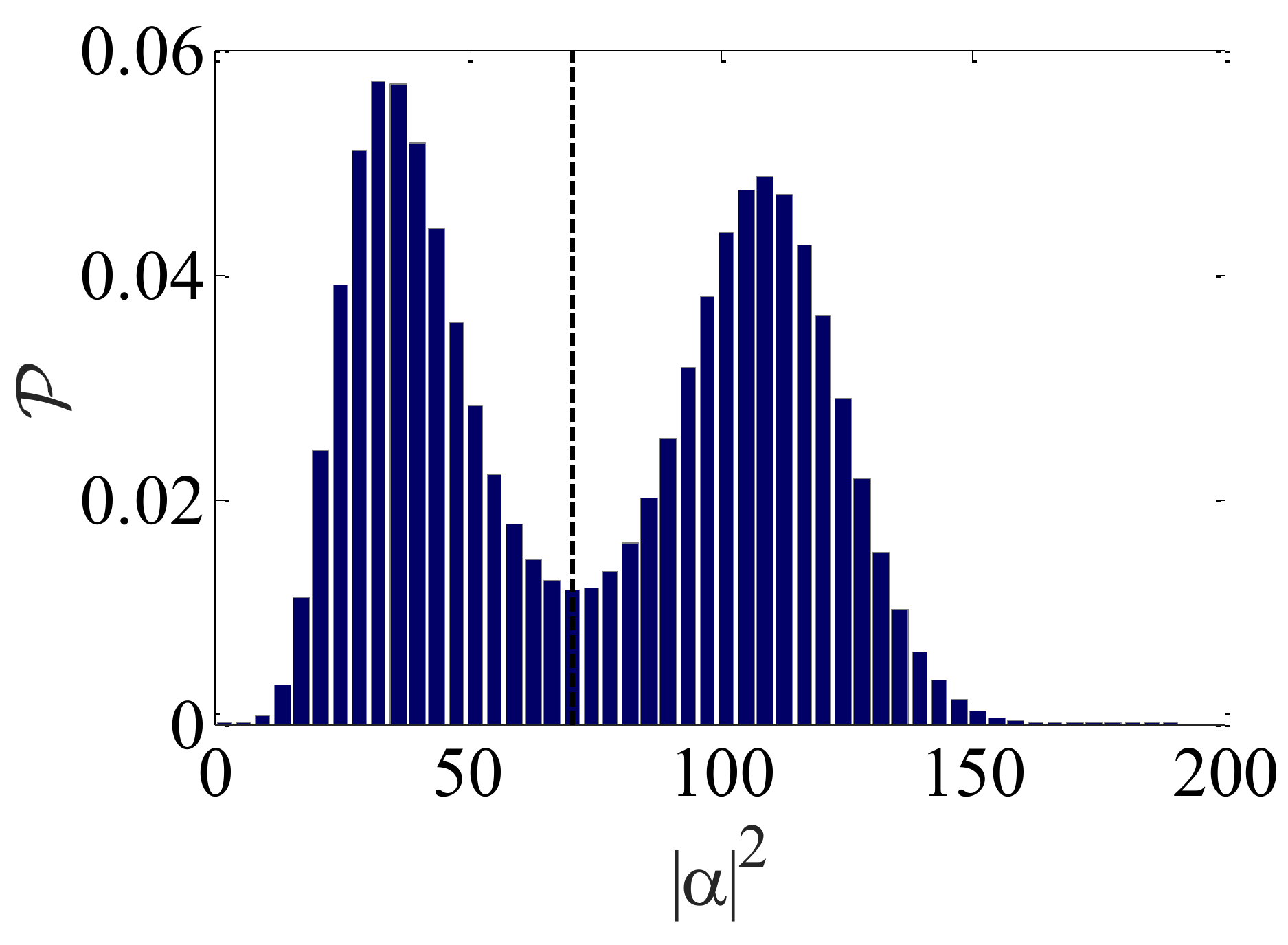}}}\caption{Probability density function  of the number of photons in the cavity, for the same parameter values  given  in Fig.~\ref{fig2}, with $\epsilon=0$. The vertical dashed line indicates a local minimum. This minimum corresponds to the threshold density $N_{th}$ above (resp. below) which the system is said to reside in state $N_{\uparrow}$ (resp. $N_{\downarrow}$).}
\label{figA}
\end{figure}

Here we explain how the threshold density $N_{th}$, separating the states $N_{\uparrow}$ and $N_{\downarrow}$, can be defined. To this end, we need a long trace of the transmitted intensity by the cavity, proportional to $|\alpha|^2$. By creating a histogram of events with different values of $|\alpha|^2$, we can calculate a probability density function of  $|\alpha|^2$   as shown in Fig.~\ref{figA}. The values of the model parameters used in the calculations of Fig.~\ref{figA} are the same as those reported in the caption of Fig.~\ref{fig2}.  The results in  Fig.~\ref{figA} show a bimodal distribution, corresponding to bistability. In-between the two peaks  there is a local minimum in $|\alpha|^2$.  This local minimum maps to the peak of the potential barrier in the DWP description. We therefore ascribe the value of $|\alpha|^2$ at this minimum to $N_{th}$.

\newpage

%

\end{document}